\documentclass[%
    aip,
    jcp,
    amsmath,amssymb,
    preprint,%
]{revtex4-1}

\usepackage[utf8]{inputenc}
\usepackage[T1]{fontenc}
\usepackage[english]{babel}
\usepackage{cleveref}
\usepackage[caption=false]{subfig}
\usepackage{graphicx}
\usepackage{physics}

\usepackage[dvipsnames]{xcolor}
\definecolor{BLUE}{rgb}{0,0,1}

\DeclareMathSizes{10}{10}{5}{4}

\newcommand*\mean[1]{\bar{#1}}

\newcommand{\ljsigma}{\sigma_\text{LJ}}
\newcommand{\Ap}{A_\text{p}}
\newcommand{\Aact}{A_\text{act}}
\newcommand{\Np}{N_\text{p}}
\newcommand{\NGE}{N_\text{GE}}

\newcommand{\Nmax}{N_\text{max}}
\newcommand{\cov}{\mathcal{C}}
\newcommand{\covge}{\cov_\text{GE}}
\newcommand{\covcmcbulk}{\cov^\text{cmc}}%_\text{bulk}}
\newcommand{\rhoddd}{\rho_\text{3D}}
\newcommand{\rhoddmax}{\rho_\text{2D}^\text{max}}
\newcommand{\rhoLG}{\rho_\text{LG}}
\newcommand{\locality}{\mathcal{L}}
\newcommand{\localityrel}{\mathcal{L}_\text{rel}}
\newcommand{\const}{\text{const}}
\newcommand{\dGm}{\Delta G_\text{m}}
\newcommand{\Tloc}{T_\text{loc}}
\newcommand{\TJloc}{\qty(T/J)_\text{loc}}
\newcommand{\Beff}{B_\text{eff}}

\newcommand{\sigmai}{\sigma_\text{i}}
\newcommand{\sigmaj}{\sigma_\text{j}}
\newcommand{\si}{s_\text{i}}
\newcommand{\sj}{s_\text{j}}
\newcommand{\Hlg}{\mathcal{H}_\text{latticegas}}
\newcommand{\Hising}{\mathcal{H}_\text{Ising}}
\newcommand{\Egain}{E_\text{gain}}
\newcommand{\Eloss}{E_\text{loss}}
\newcommand{\Tc}{T_\text{c}}
\newcommand{\TJc}{\qty(T/J)_\text{c}}

\begin{document}

\title{Frame-guided assembly from a theoretical perspective}
\author{Simon Raschke}\email{\url{simon.raschke@uni-muenster.de}}
\author{Andreas Heuer}\email{\url{andheuer@uni-muenster.de}}
\affiliation{Westfälische Wilhelms-Universität Münster, Institut für physikalische Chemie, Corrensstraße 28/30, 48149 Münster, Germany}
\date{\today}
\begin{abstract}
    The molecular self-assembly of various structures such as micelles and vesicles has been the subject of comprehensive studies. %
     Recently, a new approach to design these structures, the frame-guided assembly, has been developed to progress towards fabrics of predefined shape and size, following an initially provided frame of guiding elements.
  Here we study frame-guided assembly into a two-dimensional membrane via computer simulations, based on a single-bead coarse grained surfactant model in continuous space.
    In agreement with the experiment the assembly process already starts for surfactant concentrations below the critical micelle concentration. Furthermore, upon decreasing temperature the formation process gets more localized. Additionally, we consider a lattice gas model of the membrane plane including guiding elements where the particle concentration is varied via a chemical potential.  It behaves similar to the continuous model and additionally allows the formulation of analytical mean-field predictions which provide a  fundamental understanding of frame-guided assembly.
\end{abstract}

\maketitle

\section{Introduction}\label{sec:intro}

Our daily lives surround us with practical applications of assembled supramolecular structures. %
Their applications vary in complexity and size ranging from micelle formation using soap\cite{klevens1946critical} over neurotransmitter transport via vesicles\cite{del1956biophysical} to specific DNA-Origami\cite{rothemund2006folding} in cancer therapy\cite{zhao2012dna,zhang2014dna}.
These structures, such as complexes, enzymes and even cell membranes, are artificially constructed and serve a particular purpose, e.g. drug delivery\cite{sharma1997liposomes,tian2014doxorubicin}. %
As the desire for methods and structural shapes grew especially in surfactant and lipid systems, the increasing demand for designable building blocks \cite{bayburt2002self,zhang2017design} and especially DNA scaffolds\cite{yang2016self,franquelim2018membrane} followed. %
In \citeyear{dong2014frame} \citet{dong2014frame} published a new concept with the goal of generating vesicles of programmable size and shape based on a predefined frame, namely, the frame-guided assembly. %
The basic frame was generated by single strand DNA, of which many were attached onto the surface of a gold nanoparticle. %
Via DNA-hybridization a second, complementary DNA strand was bound to the first one and extended the strand by a head group. %
These DNA strands and their respective head groups form a frame-guide around the gold nanoparticle. %
The head group was designed in a way to allow for free monomers in bulk to be accumulated between the head groups of these guiding elements. %
As a prerequisite for the frame-guided assembly it is essential to chose a monomer concentration below the point of self aggregation. %
This implies, that the frame-guided assembly facilitates aggregation below the critical micelle concentration (cmc) of the monomers. %
By shaping the surface on which the single strand DNA is attached, the authors were able to generate many shapes of vesicles resembling the structure of the underlying base of the frame-guide. %
This building block yields a high flexibility and was also be combined with DNA-origami scaffolds\cite{dong2017cuboid} to create even more complex structures. %

In computer simulations, it is of great importance to reach time scales that are long enough to observe the phenomena of interest. %
This often is achieved by the usage of coarse-graining methods that reduce the complexity of the algorithm are used. %
Coarse graining is a widely used method in biochemical computer simulations of membranes\cite{shelley2001simulations,lenz2005,marrink2007,hakobyan2013phase}, nucleic acids\cite{sim2012modeling,maciejczyk2010coarse} and proteins\cite{baaden2013coarse,poulain2008insights}. %
The unnecessary complexity is hidden and one is able to study effects and dynamics in systems, that would otherwise be beyond the reach of contemporary computational capabilities. %

In this work, we take a detailed look into the mechanism of the frame-guided assembly method with the use of a simplistic, highly coarse-grained theoretical model and investigate the assembly of planar frame-guided structures. %
In order achieve the long time scales that are needed due to the slowdown in dynamics around cmc we take a recently developed single bead surfactant model \cite{raschke2019non} and extend it in a way to allow us to study this new assembly strategy. %
Our goal is threefold. Firstly, for that coarse-grained surfactant model we characterize the frame-guided assembly in dependence of the relevant parameters such as the density of guiding elements. Secondly, we present a microscopic picture of the underlying processes of the assembly process. Thirdly, in parallel we study a  2D lattice gas model. Beyond analogous simulations as for the surfactant model it is also possible to formulate analytical predictions based on a mean-field approximation of the effect of the guiding elements. They are compared with the outcome of the simulations and  can explain some of the key numerical results of the frame-guided assembly. Thus, the lattice gas model may serve as a kind of theoretical reference which allows us to gain additional information about the underlying behavior, e.g., with respect to the relevance of a critical temperature.

\section{Models and Methods}\label{sec:method}

\subsection{Definition continuous model}

Here, we present an application of the coarse grained model by \citet{raschke2019non}, which was developed for modelling micelle formation of surfactants in bulk. %
Due to the high flexibility of the model, it can be mapped onto various types of surfactants. %
Molecules are modelled as point-like particles with an orientation vector, which interact via a modified Lennard-Jones(12,6) potential. %
Via the orientation vector an anisotropy was introduced into the interaction. %
This made it possible to fit the model to critical packing parameters\cite{israelachvili1976theory} of various surfactant systems and carry out simulations. %
By adjusting an optimum angle of interaction \(\gamma\), the anisotropy was controlled so that structures of different size and shape could be sampled. %
In principle the application ranges from micelles to inverted micelles and is also suitable for planar structures like membranes. %

In this work, the model is extended upon for the application of the frame-guided assembly strategy. %
A frame-guide is an object of multiple structure defining elements, which facilitate cluster growth in the predefined frame. %
The structure defining elements are called the guiding elements. %
These elements can be placed in arbitrary formations of which vesicle\cite{dong2015using,dong2015frame} and planar\cite{zhou2016precisely} shapes were discussed in literature. %
Therefore,  we introduce a new particle type, the guiding element, into the model. %
We define these guiding elements to interact with the free particles in the same way free particles interact with each other.  The 3D density of the free particles is denoted \(\rhoddd\).
Guiding elements, however, are constrained in their translational and rotational degrees of freedom and are bound loosely to their initial position and orientation. %
The magnitude of these constraints are a compromise between the rigidity of the frame-guide and the ability to adapt to free particles moving in and out of the frame-guide. %

\begin{figure}
    \centering
    \subfloat[\label{sfig:fga_setup3d}]{\includegraphics[width=.44\linewidth]{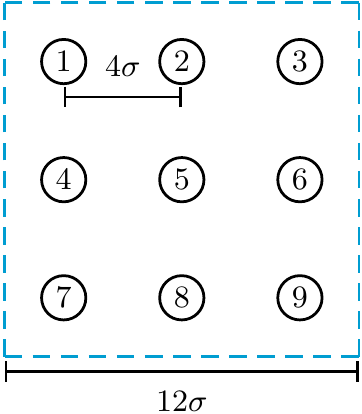}}\hfill
    \subfloat[\label{sfig:fga_setup2d}]{\includegraphics[width=.52\linewidth]{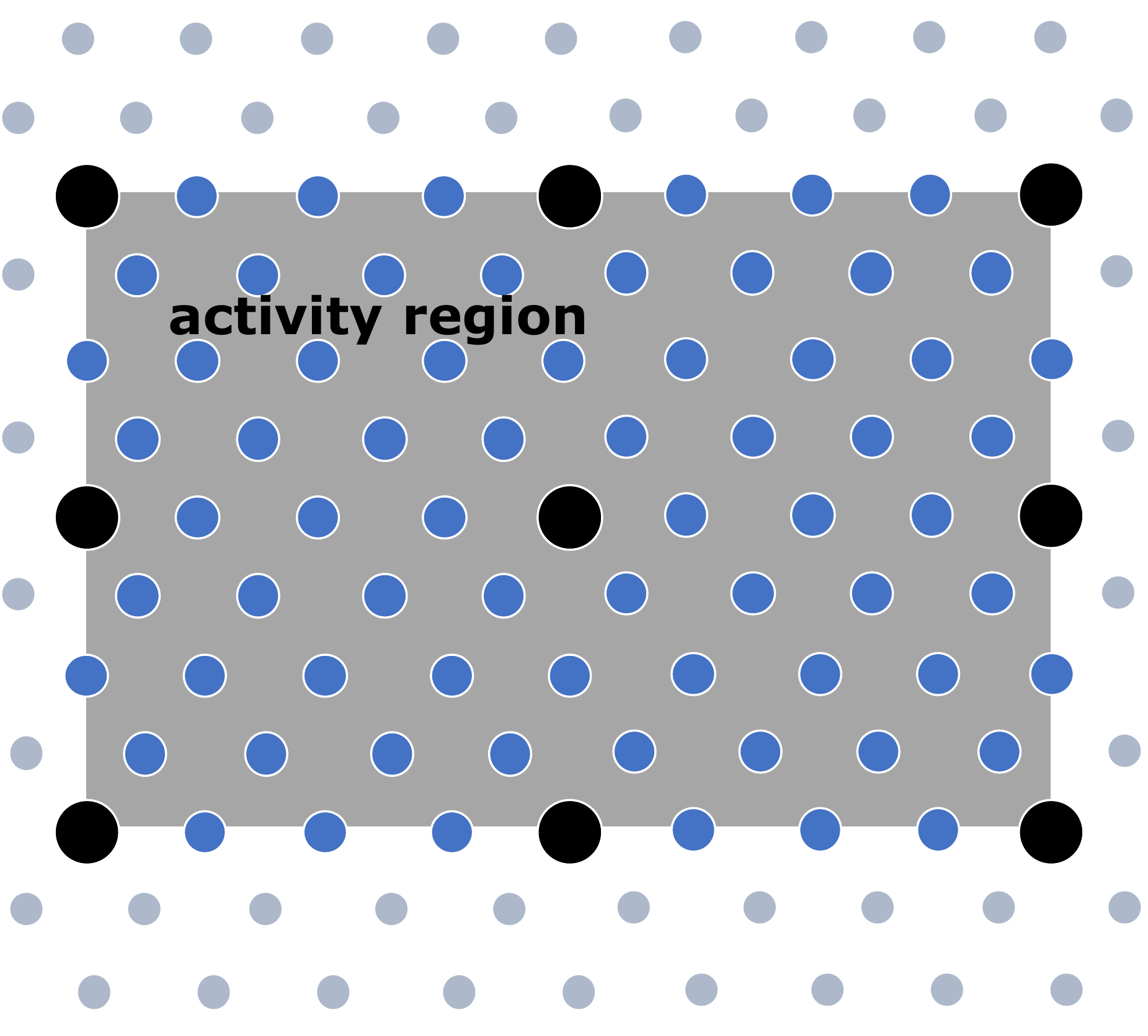}}
    \caption{Construction of a planar frame-guide with \(\NGE=9\) guiding elements. %
        The dashed line denoted setup plane marks the virtual plane with an edge length of \(12\) Lennard-Jones units \(\ljsigma\) in which the setup takes place. %
        The grey area in between the guiding elements is the activity region. %
        The setup is shown for \protect\subref{sfig:fga_setup3d} the continuous model and \protect\subref{sfig:fga_setup2d} the lattice gas model. %
    }
    \label{fig:fga_setup}
\end{figure}

The setup of the planar frame-guide was realized via placing guiding elements on a grid with the edge length of \(12\ljsigma\) in Lennard Jones units in the center of the simulation box. %
The limits of the box far exceed the size of the frame-guide in avoidance of large clusters spanning the complete box and thereby becoming artificially very stable due to not experiencing border effects when a cluster stabilizes itself across the periodic boundary conditions. %
On the grid, guiding elements were placed equidistant so that individual particles had a distance of the edge length divided by the number of guiding elements per dimension \(12\ljsigma / \NGE \), where \(\NGE\) is  the total number of guiding elements. %
\Cref{sfig:fga_setup3d} exemplarily shows the setup for \(\NGE{=}9\). %
The activity region is spanned by the outer guiding elements. %
It has the area \(\Aact=144 \sigma^2 \qty(\sqrt{\NGE} - 1)^2/\NGE\). %

In analogy to previous work on this model \cite{raschke2019non} simulations were carried out via the Metropolis\cite{metropolis1953equation} Monte Carlo\cite{metropolis1949monte} sampling method in the NVT ensemble. %
Frame-guided systems were set up with an optimum angle of \(\gamma{=}0^\circ\) in order to generate planar structures. %
Guiding elements were set up in the center of the box and free particles were distributed uniformly in a random way across the simulation box. %
The exact number of free particles is adjusted to match a predefined particle density \(\rhoddd\). %
All production simulation runs sampled \(4\cdot10^7\) translational and rotational Monte Carlo trial moves each per particle in the system. %
The step width of a Monte Carlo step was fixed to a maximum of \(0.2\ljsigma\) for translation and a maximum \(0.2\;\text{rad}\) for rotation in order to allow for a time comparison between different sets of simulation parameters. %

The temperatures $T$, used in this work, are expressed in units of the Lennard-Jones energy of the pair-interaction potential.

\subsection{Coverage continuous model}

For a given configuration of particles within the plane, spanned by the guiding elements, we want to calculate the coverage. %
For the number of particles in the activity region there are two contributions. %
We start with the effective number of guiding elements. %
Here we have to take into account that guiding elements at the edge only contribute with 50\% and those at the corner with 25\%. %
A straightforward yields for their effective number $\qty(\sqrt{\NGE}{-}1)^2$. %
A second contribution is the number of free particles $\Np$. %

The sum of these values has to be related to the maximum possible number of particles in the activity region when all particles are arranged on a triangular lattice. Then the area \(\Ap\), occupied by a single particle, is given by the area of a regular hexagon with its edge points given by the midpoints between two neighbor particles. The distance of opposite midpoints just equals the distance of two particles in the minimum of the Lennard Jones potential. Thus, the  incircle radius \(r\) is exactly half of the Lennard Jones optimum distance $ 2^{1/6}\ljsigma$ and we can write %  %
\begin{equation}
    \Ap = 2\sqrt{3} \; r^2 = 1.091\ljsigma^2\;,
\end{equation}

From this, we can derive the maximum surface density of particles as %
\begin{equation}
    \rhoddmax = \frac{1}{\Ap} = \frac{1}{1.091\ljsigma^2} = 0.9166\; \ljsigma^{-2}\;.
\end{equation}

Now we are in a position to express the plane coverage \(\cov\) as %
\begin{equation}
    \cov = \frac{\Np{+}\qty(\sqrt{\NGE}{-}1)^2}{\Aact\rhoddmax}\;.
\end{equation}
For \(\Np{=}0\) the coverage is the coverage due to the presence of the guiding elements \(\covge=\cov\qty(\Np{=}0)\). %

\subsection{Lattice gas model}

We consider a triangular lattice with $24 \cdot 24$ sites and periodic boundary conditions in 2D. %
As shown in \cref{sfig:fga_setup2d} we choose, in close analogy to the particle-based simulations, a local arrangement of \(\NGE\) guiding elements such that the number of rows/columns one has to add from one guiding element to the next amounts to $12/\sqrt{\NGE}$. %
If two adjacent sites are occupied the interaction energy is $J$.
The Boltzmann constant $k_\text{B}$ will be set to unity.%

During our Metropolis Monte Carlo procedure we attempt to move a lattice particle to an occupied and neighboring empty lattice site. %
The guiding elements are fixed in space. %
Beyond the rearrangement of lattice particle we also attempt to either generate a lattice particle on an empty site or remove a lattice particle from an occupied site. %
For the generation of a lattice particle one calculates the (possible) gain in energy $\Egain{<}0$ and adds the constant value $\mu {>} 0$, representing the negative chemical potential in the dilute limit. %
The acceptance is checked with a Metropolis Monte Carlo criterion. %
In analogy, the possible removal of a particle is realized by subtracting $\mu$ from the (possible) loss in energy $\Eloss{>}0$ and then taking the same criterion. %
In case of no interactions among the particles, i.e. $J{=}0$, the resulting equilibrium concentration is %
\begin{equation}
    \rhoLG = \frac{1}{1 + e^{\mu/T}}\;. %
    \label{eq:rhoLG}
\end{equation}
The relation to the bulk density \(\rhoddd\) of the continuous model will be discussed further below. %

When starting the simulations we have started with an lattice without free particles. In all cases we have averaged over at least 100 independent runs, each with $10^6$  Monte Carlo steps.

\subsection{Mapping to Ising model}

There is a straightforward mapping from the lattice gas model to the Ising model. %
The Hamilton of the lattice gas model reads %
\begin{equation}
\Hlg = -\frac{J}{2}\sum_{\langle \text{i,j} \rangle} \sigmai \sigmaj + \mu \sum_\text{i} \sigmai. %
\end{equation}
where the first sum for a given index $i$ runs over all neighbors $j$. %
Here $\sigmai$ expresses whether site $i$ is populated ($\sigmai {=}1$) or not ($\sigmai {=}0$). %
The mapping to the Ising model involves the transformation to spin variables $\si {=} 2\sigmai {-} 1$. %
This yields for a triangular lattice %
\begin{equation}
\label{eq:Hising}
\Hising = -\frac{J}{8}\sum_{\langle \text{i,j} \rangle} \si \sj +\qty( \frac{1}{2} \mu -\frac{3}{2} J )  \sum_\text{i} \si + \text{const}\;. %
\end{equation}

Thus, naturally the thermodynamic properties of the Ising model directly translate to those of the lattice gas model. %
In particular, the critical temperature of the lattice gas model $\Tc$ is one fourth the critical temperature of the triangular Ising model, i.e. $\qty(T/J)_\text{c} =3.6403 /4  \approx 0.91$ \cite{zhi2009critical} The inverse reads $\qty(J/T)_\text{c} \approx 1.10$.
Furthermore, from \cref{eq:Hising} the effective magnetic field can be identified as
\begin{equation}
   \Beff = (1/2)(\mu - 3J).
\end{equation}
Thus, both the chemical potential and the interaction strength contribute to $\Beff$. %
At cmc the system needs to be symmetric with respect to the spin direction since the lattice gas model with high coverage has the same free energy as the system with low coverage. %
As a consequence at cmc one exactly has $\Beff{=}0$. %

In the Ising representation a guiding element can be described as a spin which is in a permanent up state $\si {=} 1$. %
Thus, a neighbor j of a guiding element experiences an additional energy contribution, related to the presence of the guiding element, given by  $-(J/4)(1 - \langle s \rangle ) \sj$. Here $\langle s \rangle$ denotes the average magnetization. Since we are mainly interested in the behaviour around cmc, one has $\langle s \rangle=0$.
If the coverage of guiding elements is denoted \(\covge\) and taking into account that a guiding element has 6 neighbors one gets %
\begin{equation}
   \Beff = (1/2)(\mu - 3J - 3J \covge).
\end{equation}
This relation implies that the impact of a guiding element is equally distributed over all sites, which corresponds to a mean-field picture.
Since at cmc one has $\Beff{=}0$ one can calculate the required chemical potential via
\begin{equation}
 \mu =3J \cdot (1+\covge).
 \label{eq:mutheo}
\end{equation}
Together with \cref{eq:rhoLG} this yields (using $\mu/T \gg 1$ which will always be the case)
\begin{equation}
\text{cmc} = \exp[-3(J/T)] \exp[-3(J/T)\covge].
\label{eq:cmc_exact}
\end{equation}

For the coverage of the lattice gas model we just count the occupied sites in the activity region and compare them with the total number of sites. Again, the effective number of guiding elements, to be taken into account in the activity region, is $(\sqrt{\NGE}-1)^2$ .

\subsection{Quality of mean-field approximation}

\begin{figure}
    \centering
    \subfloat[\(J T^{-1}{=}1.2\).\label{sfig:mu_theory_actual_J12}]{\includegraphics[width=.5\linewidth]{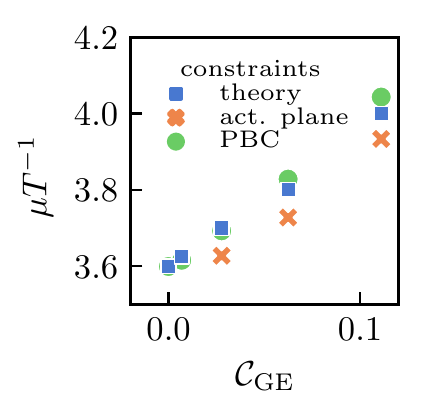}}\hfill
    \subfloat[\(J T^{-1}{=}1.8\).\label{sfig:mu_theory_actual_J18}]{\includegraphics[width=.5\linewidth]{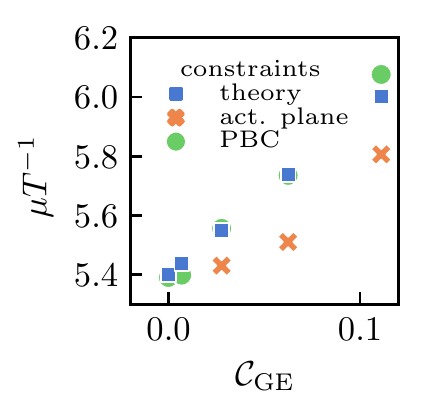}}
    \caption{Chemical potential divided by the temperature \(\mu T^{-1}\) as a function of the guiding element coverage \(\covge\) under periodic boundary conditions (PBC), in the activity region and for the mean-field theory for \protect\subref{sfig:mu_theory_actual_J12} \(J T^{-1}{=}1.2\) and \protect\subref{sfig:mu_theory_actual_J18} \(J T^{-1}{=}1.8\).  For the case of PBC, $\NGE$ is chosen to be $0^2, 1^2, ..., 4^2$ whereas for the analysis of the activity region data are shown for $2^2, 3^2, 4^2$. %
    }
    \label{fig:mu_theory_actual}
\end{figure}

We checked the quality of the mean-field approximation for two different temperatures, $J/T{=}1.2$ and $J/T{=}1.8$. Specifically, we determined via systematic variation of the chemical potential, based on interval bisectioning, when the total system is at cmc, i.e. on average half of the available sites are populated. %
This value can be compared with the mean-field result in \cref{eq:mutheo}.
Two different scenarios were evaluated. %
First, we used a system with $12 \cdot 12$ sites together with periodic boundary conditions. %
In this way the boundary effects at the active plane are removed. %
As seen in \cref{fig:mu_theory_actual}, the agreement between simulation and mean-field approximation is nearly perfect for both temperatures. %
In the second scenario we just considered the activity region in the larger system. %
For extrapolation to low coverage \(\covge\) the results, obtained from analysis of the activity region, agree with the simulations, using periodic boundary conditions. This is a necessary consequence of the condition $\Beff=0$ at cmc. Interestingly, the dependence on  \(\covge\) is somewhat weaker which holds in particular for the lower temperature. Thus, application of \cref{eq:cmc_exact} should be particularly suited for higher temperatures.

\section{Results}\label{sec:Results}

\begin{figure*}
    \centering
    \subfloat[Setup.\label{sfig:snap_setup}]{
        \begin{tabular}{@{}c@{}}
            {\frame{\includegraphics[width=.23\linewidth, height=4cm]{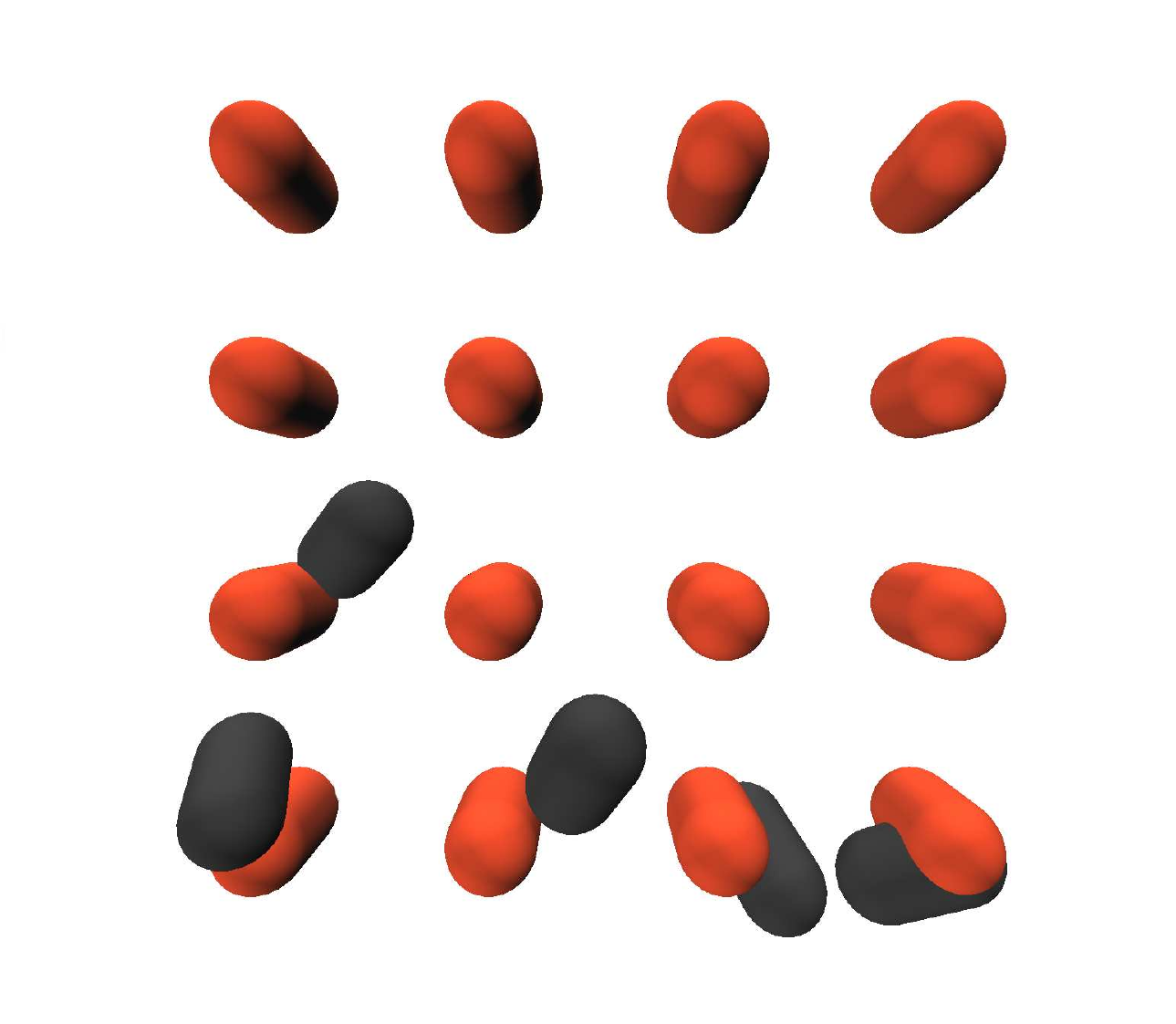}}}
        \end{tabular}
    }\hfill
    \subfloat[Onset of agglomeration.\label{sfig:snap_first}]{
        \begin{tabular}{@{}c@{}}
            {\frame{\includegraphics[width=.23\linewidth, height=4cm]{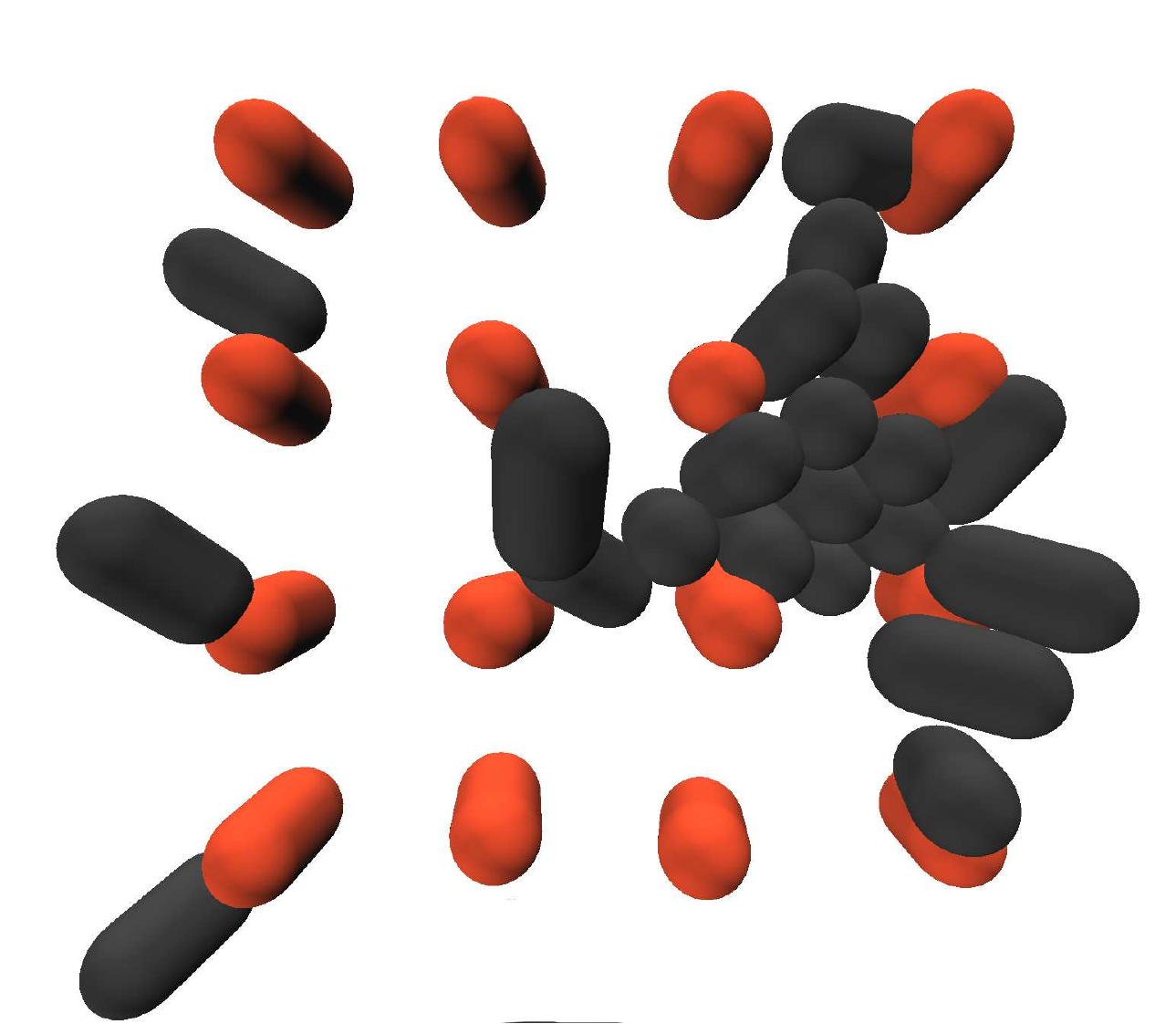}}}\\[2ex]
            {\frame{\includegraphics[width=.23\linewidth, height=4cm]{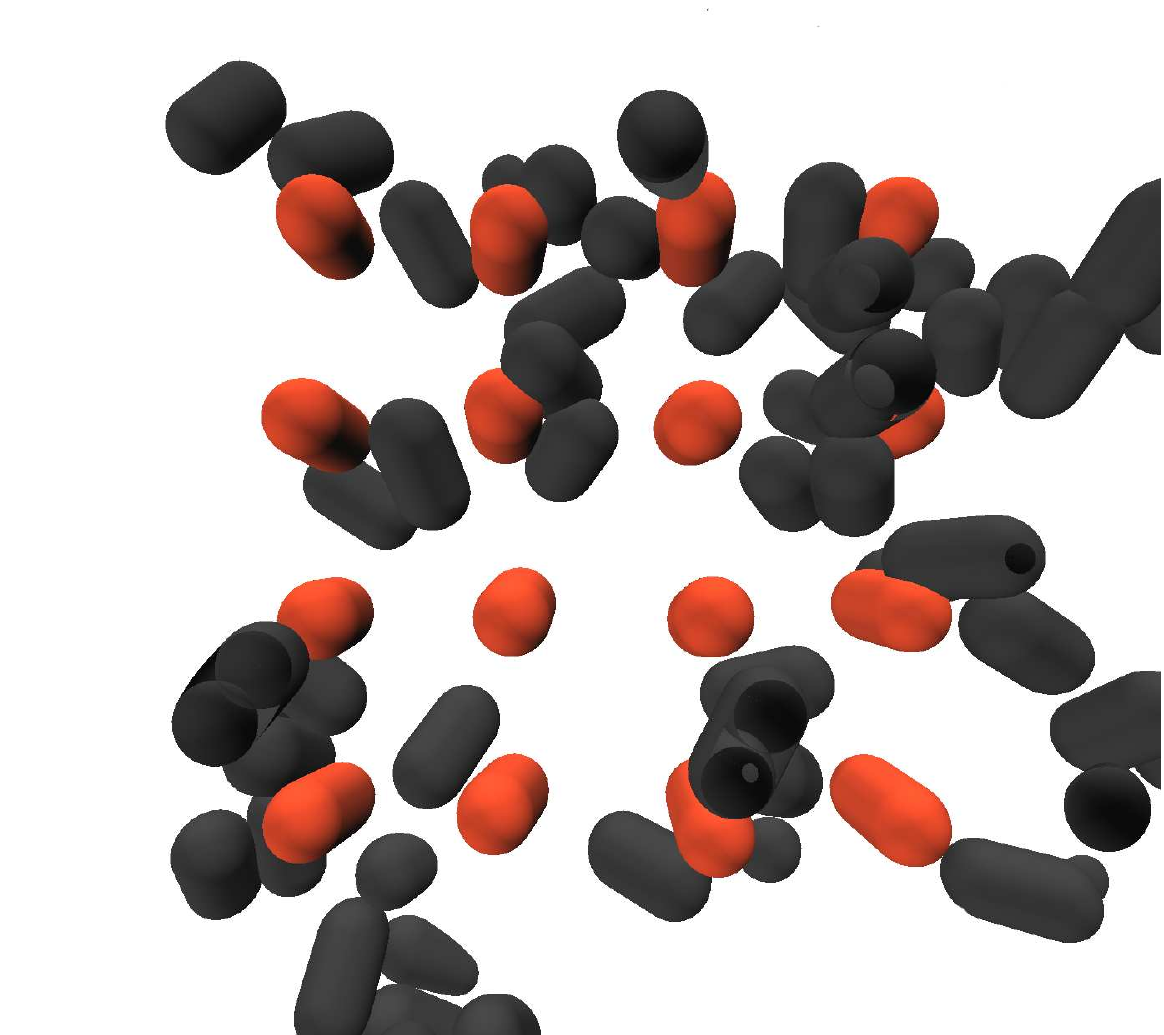}}}
        \end{tabular}
    }\hfill
    \subfloat[Growing structures.\label{sfig:snap_continued}]{
        \begin{tabular}{@{}c@{}}
            {\frame{\includegraphics[width=.23\linewidth, height=4cm]{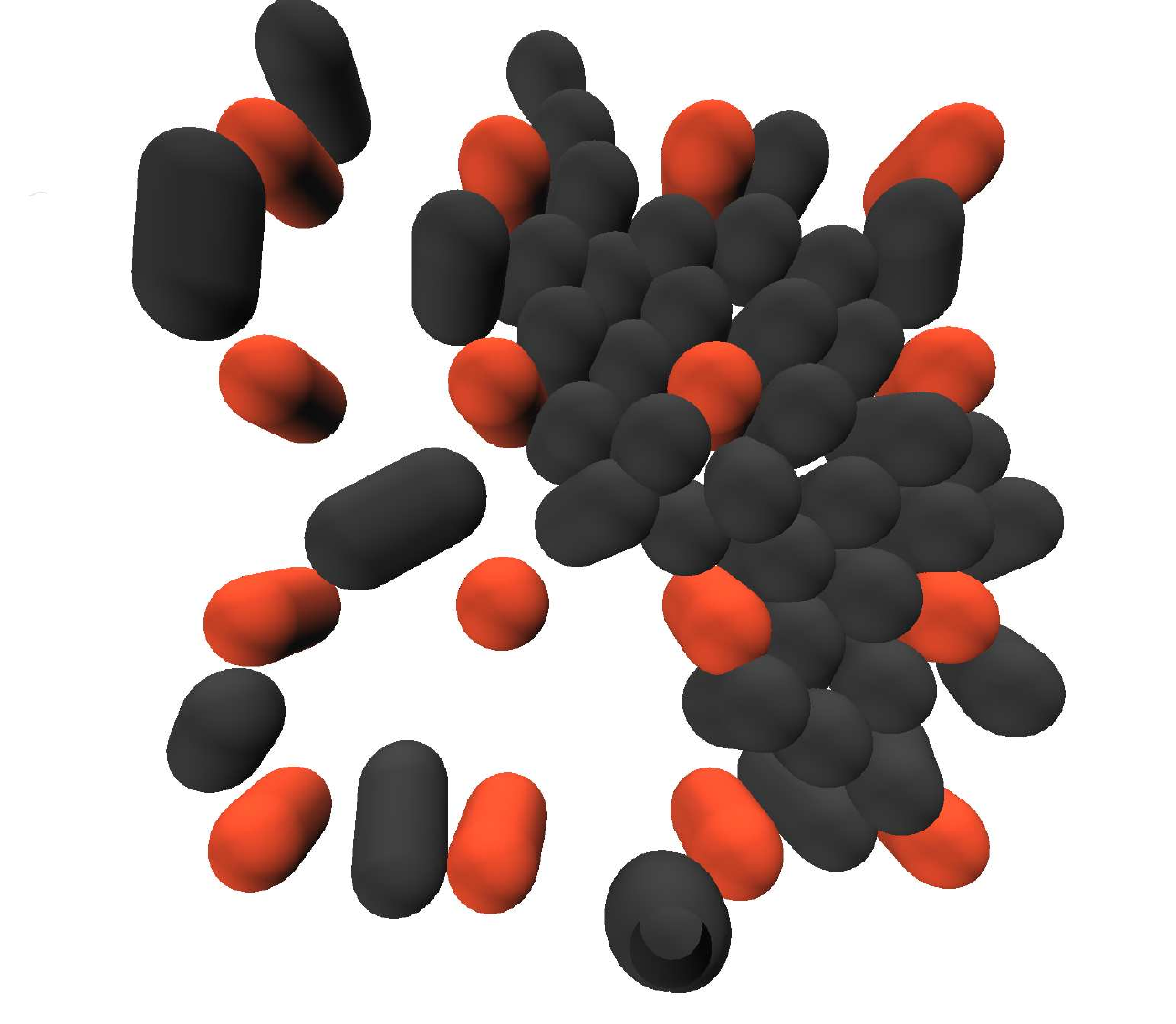}}}\\[2ex]
            {\frame{\includegraphics[width=.23\linewidth, height=4cm]{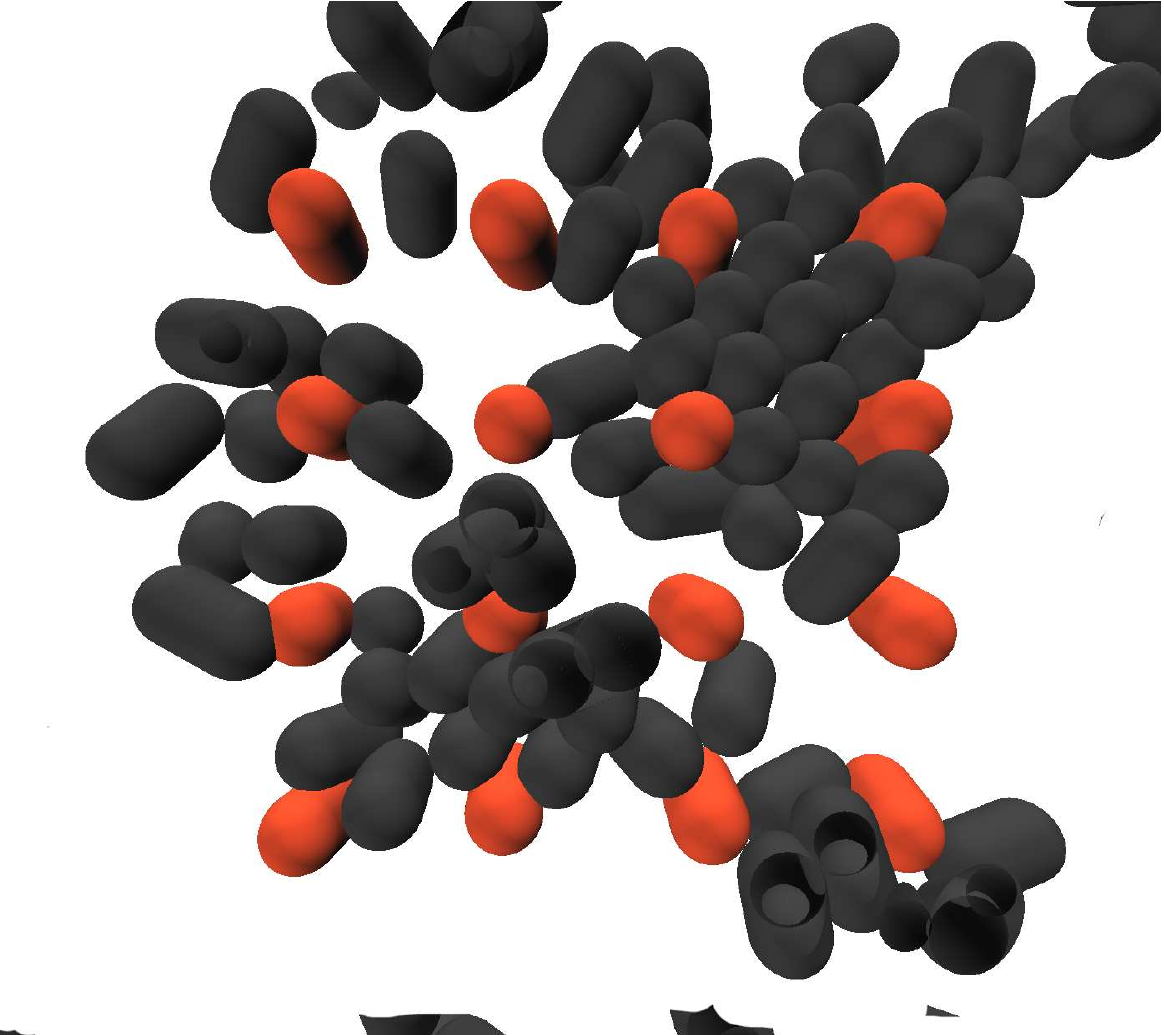}}}
        \end{tabular}
    }\hfill
    \subfloat[Fully assembled plane.\label{sfig:snap_complete}]{
        \begin{tabular}{@{}c@{}}
            {\frame{\includegraphics[width=.23\linewidth, height=4cm]{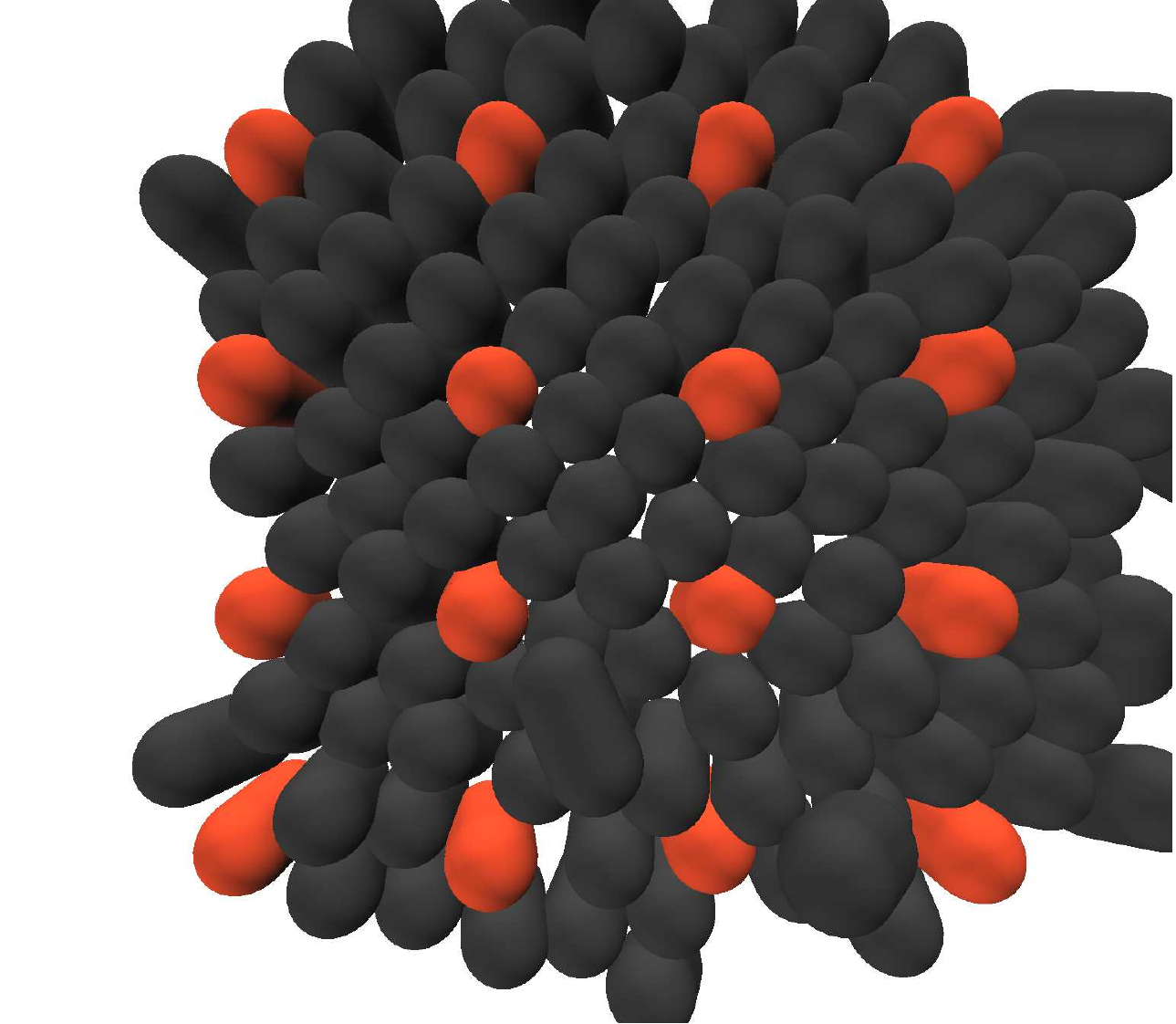}}}
        \end{tabular}
    }
    \caption{%
        Extracts from snapshots of the assembly process of a planar frame-guided system with 16 guiding elements \protect\subref{sfig:snap_setup} in its initial setup, \protect\subref{sfig:snap_first} while the first agglomeration is formed, \protect\subref{sfig:snap_continued} as the cluster grows and, \protect\subref{sfig:snap_complete} as a completely assembled plane. %
        The above row shots the assembly process as it takes place in a localized fashion from a single nucleation site. %
        The bottom row presents a less localized more general assembly process. %
        Particles are represented as sticks, where the stick denotes the center of mass between the endpoints and the orientation vector from endpoint to endpoint. %
        Guiding elements are depicted in red, while free particles are shown in black. %
    }
    \label{fig:assembly_snapshots}
\end{figure*}

\subsection{Qualitative behavior}

An example of cluster formation is shown  in \cref{fig:assembly_snapshots} at different stages of the assembly process of a planar frame-guided cluster for the continuous model. The concentration is chosen such that in equilibrium the plane is (nearly) fully covered.
Prior to the assembly process, free particles interact with guiding elements randomly and dissociate after some simulation steps. %
Once a certain amount of free particles is located between guiding elements, these particles are stabilized in the guiding element structure as one observes in \cref{sfig:snap_first}. %
The assembled particles function as a nucleation site for further cluster assembly, which is facilitated further via the next guiding elements in proximity. %
This is depicted in \cref{sfig:snap_connected}, where the frame-guide is approximately half-filled with free particles. %
Note that the assembly process can be more localized as seen in the upper figures or more delocalized as shown in the lower figures. This will be quantified further below.
For long times the activity region is fully covered by particles, extending even beyond the activity region, see \cref{sfig:snap_complete}. As expected the particles basically arrange in a triangular lattice.

\subsection{Coverage}

\begin{figure*}
    \centering%
    \subfloat[Continuous   model.\label{sfig:plane2d_coverage3d}]{\includegraphics{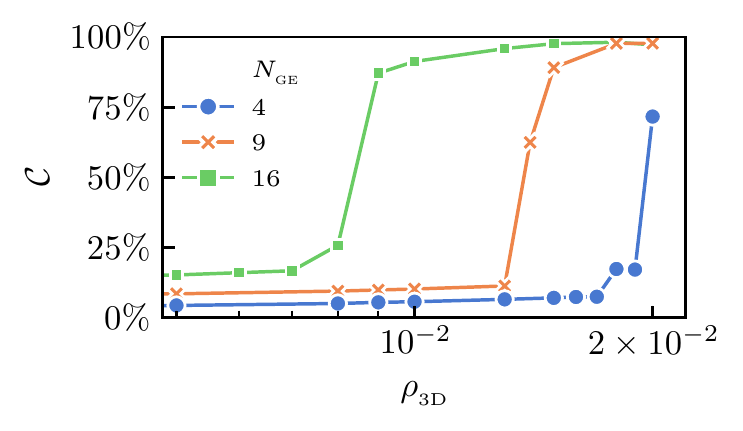}}\hspace{1cm}%
    \subfloat[Lattice gas   model.\label{sfig:plane2d_coverage2d}]   {\includegraphics{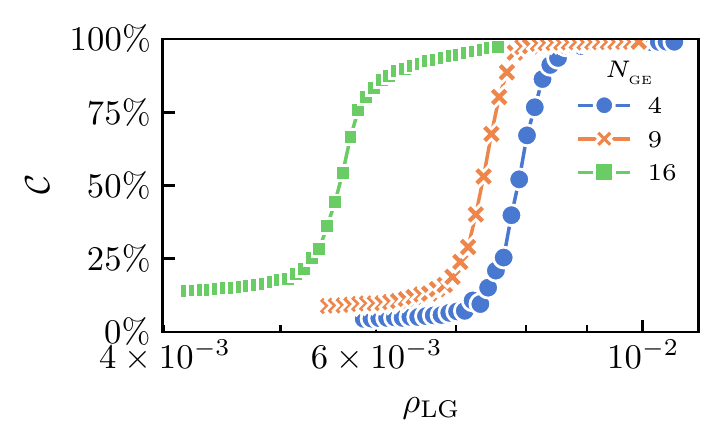}}%
    \caption{%
        Coverage \(\cov\) of the guiding element plane with particles \protect\subref{sfig:plane2d_coverage3d} as a function of the overall particle density \(\rhoddd\) at temperature \(T{=}0.23\) in the continuous 3D model and \protect\subref{sfig:plane2d_coverage2d} as a function of the density \(\rho\)  at temperature \(J{=}1.4\)  the 2D lattice model with 4, 9 and 16 guiding elements \(\NGE\). %
    }%
    \label{fig:coverage}
\end{figure*}

We start by studying the coverage \(\cov\) of the activity region which can hold values between \(\covge\) and 1. It is shown in \cref{fig:coverage} as a function of \(\rhoddd\) at various \(\NGE\). %
In general, one observes a steep increase in plane coverage at certain \(\rhoddd\) values.
One can state that the cluster formation process is very sensitive to changes in particle density. Small changes shift the system from no cluster formation to a fully populated activity region. %

Most importantly, with the addition of guiding elements to the system, cluster formation was observed below the cmc of the non-guided system, i.e. the value of cmc decreases with increasing \(\NGE\). %
This reflects the impact of the frame-guide to provide nucleation sites for free particles. Since we work below the cmc of the cluster formation without a frame-guide, cluster formation is not observed outside of the frame-guide. Thus, all processes of interest are expected to occur in the activity region. %

A fully analogous picture we obtain for the lattice gas model. Again, the presence of guiding elements reduces the density where the coverage strongly starts to increase.

\subsection{Efficiency of guiding elements}\label{subsec:efficiency}

\begin{figure}
    \centering
    \includegraphics{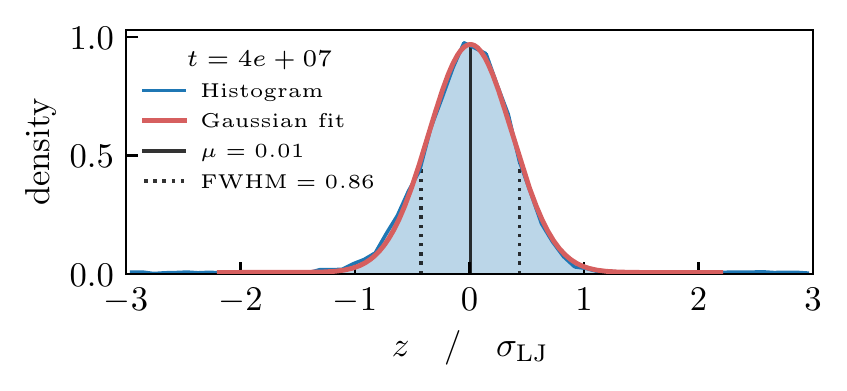}
    \caption{Probability density of particles in z-direction of the simulation box centered around the plane of the frame-guide \(z{=}0\). The histogram normalized to 1 and an ensemble average of 10 systems at \(t=4\cdot10^7\) simulation steps with a temperature of \(T{=}0.22\) and \(\NGE{=}16\) guiding elements. The red line represents the Gaussian fit of the histogram, and the dashed lines indicate the full width at half maximum FWHM of the Gaussian. }%
    \label{fig:zfit}
\end{figure}

Now the dependence of cmc on $\NGE$ is discussed in more detail.
We define cmc as the lowest particle density in the activity region at which \(50\%\) of the trajectories showed a density greater or equal to half of the maximum density \(\rhoddmax\). %
Note that cmc is a specific value of \(\rhoddd\). Thus, we need to convert \(\rhoddd\) into a coverage representing the 3D particle density in a planar 2D structure with layer thickness $d$. In this way it is possible to define cmc via the coverage criterion and to compare the resulting 2D density at cmc with the corresponding values of the lattice gas model.

The thickness of the particle layer in the frame-guide is depicted in \cref{fig:zfit} as the full width at half maximum FWHM of the gaussian fit of the \(z\)-distribution of particles in the box. %
Here, with \(\NGE{=}16\), the cluster formation showed no significant bending of the membrane. Therefore, we take the resulting value of \(d{=}\text{FWHM}{=}0.86\ljsigma\) as a  reasonable estimate of the layer thickness and will use it as a constant for all \(\NGE\) values in further calculations. We checked that in the range of relevant temperatures this value is insensitive to temperature.%

Accordingly, we express the 2D coverage  at the 3D critical micelle concentration as %
\begin{equation}
    \covcmcbulk = d \; \Ap \; \left.\rhoddd\right|_\text{cmc}\;.
\end{equation}
Here\(\left.\rhoddd\right|_\text{cmc}\) is the overall particle density at cmc. %
The term \(d \; \Ap\) expresses the volume one particle occupies in the triangular lattice. %

\begin{figure*}
    \centering%
    \subfloat[Continuous model.\label{sfig:cmc_plane_3d}]{\includegraphics{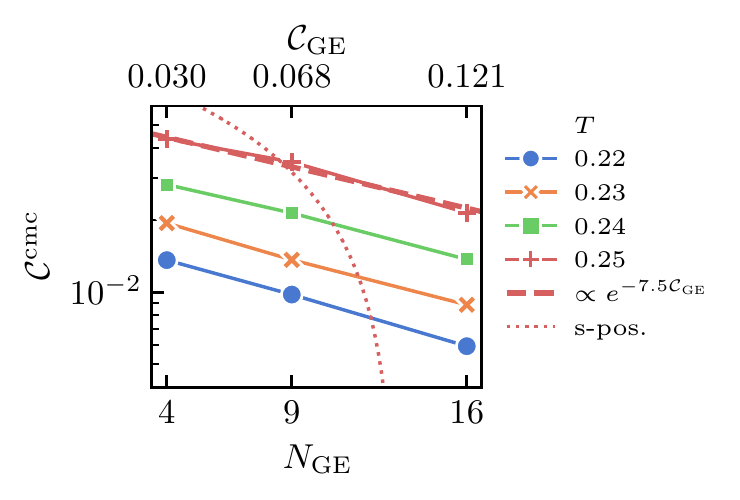}}\hspace{1cm}%
    \subfloat[Lattice gas model.\label{sfig:cmc_plane_2d}]{\includegraphics{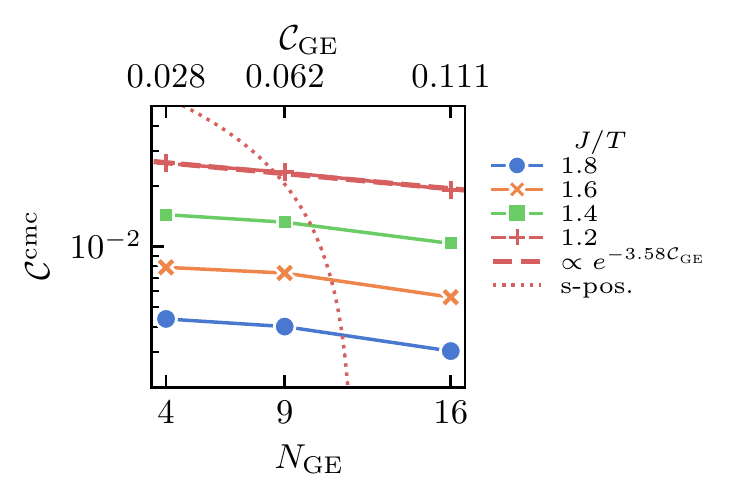}}%
    \caption{%
        Bulk coverage at the critical micelle concentration \(\covcmcbulk\) as a function of guiding elements \(\NGE\) at various temperatures \(T\). The top axis shows the complementary coverage of guiding elements \(\covge\) \protect\subref{sfig:plane2d_coverage3d} in the activity region of the continuous 3D model and \protect\subref{sfig:plane2d_coverage2d} \(\cov^\text{cmc}\) in the 2D lattice model. The broken line is the prediction of the superposition hypothesis (see text). The offset in y-direction is chosen arbitrarily.%
    }
    \label{fig:cmc_plane}
\end{figure*}

\(\covcmcbulk\) was calculated at \(T{=}0.22\), \(0.23\), \(0.24\) and \(0.25\) and \(\NGE{=}4\), \(9\) and \(16\) guiding elements as depicted in \cref{fig:cmc_plane}. %
We observed a decrease in cmc with the increase of guiding elements in the system at all sampled temperatures in agreement with the results in \cref{fig:coverage}. Furthermore, as expected, the value of cmc gets smaller at lower temperatures. The slope of the logarithm of  \(\covcmcbulk\) turns out to be approximately temperature independent.

To quantify the efficiency of guiding elements for the cluster formation process we started from the {\it superposition hypothesis}, that a certain coverage of guiding elements \(\covge\) is equivalent to the same coverage with free particles. %
In this limiting case, the sum
\begin{equation}\label{eq:superpos}
\covge + \qty(1-\covge)\;\covcmcbulk
\end{equation}
should have no dependence on the bulk density. Here \(\qty(1-\covge)\) represents the area already occupied by guiding elements and \(\covcmcbulk\) is the 2D representation of particle density at cmc in bulk. %
Since \mbox{\(\covge{\ll}1\)}, from \cref{eq:superpos} we may conclude %
\begin{equation}
    \covcmcbulk = \const - \covge%\ .
\end{equation}
The numerical relation between $\covge$ and $\covcmcbulk$ is shown in \cref{fig:cmc_plane} for different temperatures together with the prediction of the superposition hypothesis (formulated for the data at $T=0.25$). The bending results from the logarithmic representation of the y-axis.
Obviously the superposition hypothesis predicts a much stronger dependence on the number of guiding elements as compared to the actually observed dependence and, furthermore has a different functional dependence on $\NGE$.  %
We conclude, that guiding elements are capable of allowing cluster formation below the cmc-value of the non guided system, but are far less efficient in doing so when compared to a corresponding increase of the bulk density. %
This is due to the conceptual difference between frame-guided particles and free particles. %
In the latter case they represent the overall 3D concentration, i.e. the ability to feed additional particles for further growth. %
In contrast, the frame-guided particles just represent a localized 2D density.

It is also straightforward to argue why the superposition hypothesis has to fail when starting from \cref{eq:cmc_exact}.  This relation yields $\text{cmc} \approx  \exp\qty(-3J/T) {-} \qty(3J/T) \exp\qty(-3J/T) \covge $. The prefactor of $\covge $ is significantly smaller than one (since $x \exp(-x) \ll 1$ for $x \gg 1$). As mentioned above this again reflects the fact that the guiding elements just modify the chemical potential but do not change the effective density of particles which is connected with the exponential of the chemical potential.

Obviously, to a good approximation the data display an exponential dependence on $\NGE$. Indeed, whereas this was expected for the lattice model, this behavior, at least on a qualitative level, also holds for the continuous model.

\subsection{Locality of cluster formation}\label{subsec:locality}

\begin{figure*}
    \centering%
    \subfloat[Continuous model.\label{sfig:localization3d}]{\includegraphics{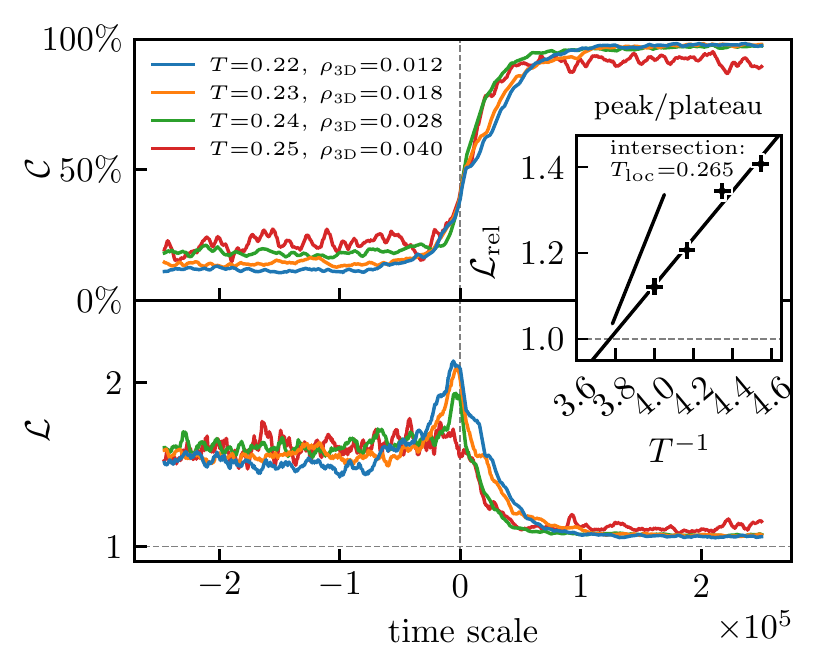}}\hfill%
    \subfloat[Lattice gas model.\label{sfig:localization2d}]   {\includegraphics{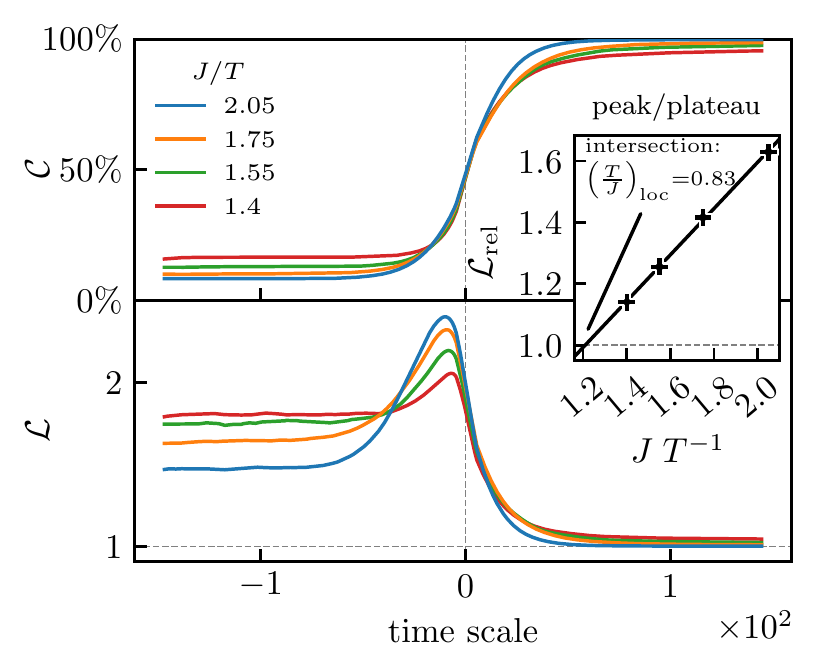}}%
    \caption{%
        Frame-guide coverage and locality of cluster formation \(\locality\) as a function of simulation time at various temperatures \(T\) and particle densities \(\rhoddd\) for 9 guiding elements above cmc for the continuous model and the lattice gas model. %
        The time scale is shifted such that  just at \(t{=}0\) a threshold of \(\cov{=}50\%\) coverage was exceeded for an individual run. %
        In \protect\subref{sfig:localization3d} every curve shows an ensemble average across multiple trajectories and a rolling mean a \(5\cdot10^3\) steps time frame. %
        The insets show the ratio of locality peak heights and the plateau values before cluster growth.
        Here, the intersection of the linear regression curves (black line) extrapolate to a temperature, where the cluster growth would occur in a delocalized way. %
    }
    \label{fig:localization}
\end{figure*}

The behavior, shown in \cref{fig:assembly_snapshots}, indicates that the complete assembly can be  preceded by localized agglomeration of the particles. %
Here we want to quantify the relevance of localized behavior. %
We took the activity region and divided it into specific domains each spanning the area between 4 guiding elements. %
In this way a grid of 9 guiding elements, to be analysed in this context, results in 4 distinct domains (see \cref{fig:assembly_snapshots}).

Within these domains the number of particles, populating each domain, was tracked. Since every guiding element is contributing to 4 domains, the number of particles per domain contain one guiding element.%
We calculated the locality as the ratio of the number of particles in the domain with the highest population\(N_\text{max}\) and the average number of particles per domain \(\mean{N}\) in the activity region %
\begin{equation}
    \locality=\frac{\Nmax}{\mean{N}}\;.
\end{equation}
When averaging over different clustering events, we shift the relative times such that at time 0 the coverage is half between minimum and maximum coverage, i.e. close to 50\%. %

Furthermore, we have chosen the density approx. 10\% above the cmc, as given in \cref{fig:cmc_plane}. %
For the lattice gas model we adjusted the chemical potential accordingly. %
This just guarantees that at some stage the system manages to form a large cluster in the activity region.
The subsequent results are very insensitive to the exact value of that increase.  %

\Cref{fig:localization} shows the results for the plane coverage in the upper half as well as the locality of cluster formation as a function of time. %
Data are presented for both the continuous model and the lattice gas model. %

For early times one has a few scattered particles in the different domains which gives rise to a time-independent value of the $\locality$, reflecting the stochastic fluctuations of particle number per domain. %
In the long-time limit every domain is nearly fully covered. By definition, this gives rise to a $\locality$ value of unity.
Of key interest is the peak of $\locality$ which at the lower temperatures appears for slightly negative time scales, i.e. shortly before the half coverage is reached. %
On this qualitative level the observations are fully analogous for the two models.

At the point where the assembly process begins we observed a steep increase in the locality of the process.
The magnitude of this effect is strongly dependent on system temperature, showing stronger localization at lower temperatures.
To characterize the height of the localization peak we divided the maximum value of $\locality$  by the average $\locality$ value before cluster formation; see the inset of \cref{sfig:localization3d}. This value is denoted as $\localityrel$.%
We observed that the vanishing of the peak, corresponding to $\localityrel$, occurs at a well-defined temperature $\Tloc= 0.265$ for the continuous model and $\TJloc = 0.83$ for the lattice gas model. %
This temperature is slightly below the critical temperature $\TJc = 0.91$.
In the next section a physical picture is presented which relates these observations to the phenomenology of the Ising model.

For the subsequent discussion we explored that $\Tloc$ can be used as a characteristic energy scale which, in principle, would allow to compare both models more quantitatively.

\subsection{Gibbs free energy}\label{subsec:gibbs}

\begin{figure*}
    \centering%
    \subfloat[Continuous model.\label{sfig:deltaGfit3d}] {\includegraphics{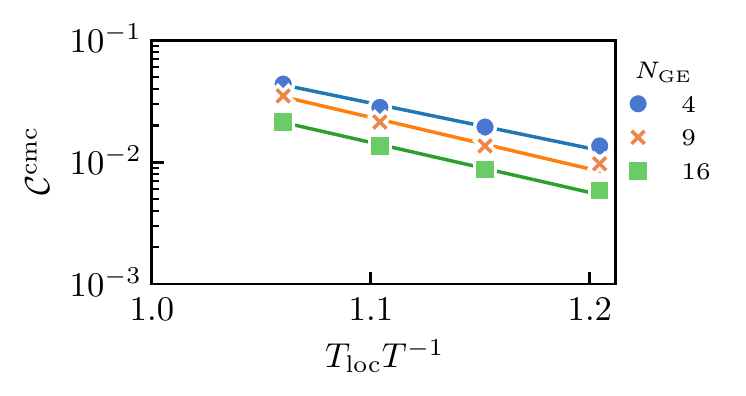}}%
    \subfloat[Lattice gas model.\label{sfig:deltaGfit2d}]{\includegraphics{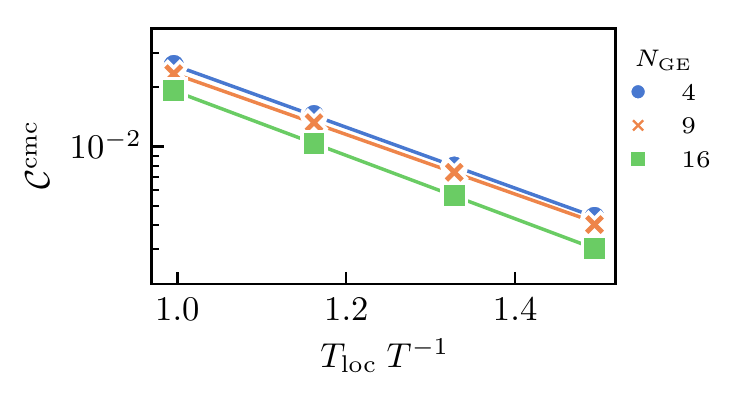}}%
    \caption{%
        Critical micelle concentration \(\covcmcbulk\) as a function of inverse temperature with respect to temperature of localization \(\Tloc\;T^{-1}\) for various number of guiding elements \(\NGE\) in the \protect\subref{sfig:deltaGfit3d} continuous model and \protect\subref{sfig:deltaGfit2d} the lattice model as comparison. %
        Solid lines indicate a fit in the form of \mbox{\(\covcmcbulk \propto \exp\qty(\frac{\dGm}{RT})\)}. %
        Temperatures are scaled to the temperature of localization \(\Tloc\). %
    }%
    \label{fig:detlaGfit}
\end{figure*}

Finally, we analyse how the value of cmc depends on temperature for fixed number of guiding elements. The results are shown in \cref{fig:detlaGfit}, where temperatures are scaled by \(\Tloc \).

\begin{figure}
    \centering
    \includegraphics{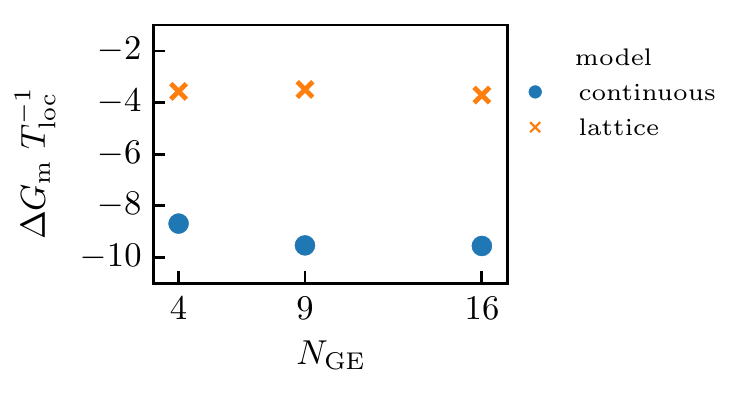}
    \caption{%
        Gibbs free energy change of micellization with respect to the temperature of localization \(\dGm\;\Tloc^{-1}\), derived from the exponential fits in \cref{fig:detlaGfit}, as a function of guiding elements \(\NGE\) comparing the continuous 3D and the lattice 2D models.%
    }%
    \label{fig:deltaG}
\end{figure}

To analyse the temperature dependence one may start with the law of mass-action regarding micellization as described by \citet{olesen2015determination}. It turns out that the  Gibbs free energy change of micellization can be expressed as %
\begin{equation}
    \text{cmc} = a_T e^{\frac{\dGm}{RT}}
\end{equation}
with a prefactor $a_T=1$.
This is also compatible with \cref{eq:cmc_exact}.
In order to reduce the impact of the presence of guiding elements our later discussion will mainly focus on the case of lowest number of guiding elements, i.e.  $\NGE=4$.  For reasons, to be discussed below, $a_T$ is treated as an additional adjustable parameter when performing the corresponding Arrhenius fit.

The results for the normalized value \(\dGm/\Tloc \)  are shown in \cref{fig:deltaG}. %
One notices the significant difference in \(\dGm\) in this scaled representation. For $\NGE=4$ we obtain  \(\dGm/\Tloc = -8.55\) for the continuous model and \(\dGm/\Tloc = -3.57\) for the lattice gas model. The corresponding values of $a_T$ are 360 for the continuous model and 0.96 for the lattice gas model. Thus, the continuous model displays a major deviation from $a_T=1$.

\section*{Discussion and Summary}

We observed that the presence of guiding elements has two main effects. %
First, the assembly process may already start below the cmc value which one would find without guiding elements. %
Second, the presence of guiding elements determines the spatial region where the assembly takes place. %
In this work, we mainly concentrated on the theoretical understanding of the variation of the cmc value upon addition of guiding elements as well as the nature of the assembly process.

Additionally, we performed simulations of a lattice gas model. We observed an increased locality index $\locality$ with decreasing temperature. %
It allowed us to define the localization temperature $\Tloc$. %
For the Ising model the value of $\Tloc $ is 10\% lower than the theoretical critical temperature. Indeed, one expects that for a finite system the observed crossover behavior occurs at slightly lower temperatures due to the additional fluctuations. Thus, it is reasonable to relate $\Tloc$ to the critical temperature. This relation is not a pure coincidence but is consistent with the observed phenomenology of the assembly process. For the Ising model it is known that the correlation length of the fluctuations is very large close to the critical temperature. Since we are dealing with concentrations close to cmc, corresponding to  $B_{eff} \approx 0$, indeed we expect a highly correlated and thus delocalized behavior.  With decreasing temperature the correlation length decreases. Thus, one expects the assembly process to be more localized as also seen from the behavior of the locality index. A second key prediction for the lattice model is the exponential dependence of cmc on the density of guiding elements. To a good approximation, this was seen for the lattice model. In particular, the alternative scenario, based on the superposition hypothesis, could be discarded.  This supports the notion that the guiding elements serve as an additional field which shift the chemical potential as derived within a mean-field approximation.

Importantly, both key results, namely the localization properties and the exponential dependence of cmc on the density of guiding elements, was also observed for the continuous model. This strongly suggests that the lattice model captures essential properties of the particle-based continuous model. Thus, via this analogy the simulations of the continuous model have gained an additional theoretical basis. In particular, the values of $\Tloc$ may serve as a measure of the respective energy scale.

From a more quantitative perspective, however, deviations between both models are present. (1) The value of $a_T$, used for fitting the temperature dependence of cmc, significantly deviates from unity for the continuous model. In contrast, the lattice model fulfills the expectation $a_T \approx 1$. This is basically equivalent to the observation that the Gibbs free energy relative to $\Tloc$, i.e.  \(\dGm/\Tloc \) is higher for the continuous model. (2) The localization effects display a stronger temperature dependence for the continuous model.  (3) The sensitivity of cmc on the number of guiding elements, expressed by the dimensionless slope $a_{GE}$, is  higher for the continuous case although in both cases temperatures close $\Tloc$ are taken, i.e. one works in comparable regions of parameter space.

Here we argue that on a qualitative level these deviations may be related to the additional complexity of the particle-based continuous model as compared to the simple lattice model.
First, the interaction between two particles in the continuous model strongly depends on the orientation. By construction of the potential, only for parallel alignment in optimum distance the full Lennard-Jones attraction is possible. When just considering a pair of particles, due to entropy effects the effective interaction strength $J$ is lower than the maximum possible interaction strength. This effect becomes smaller for lower temperatures. Naturally, entropy effects are less pronounced in this limit. This was already shown in \cite{raschke2019non}. Furthermore, at lower temperatures the growth mode is more localized as discussed above. A locally higher density naturally gives rise to a stronger parallel alignment as shown in \cref{fig:assembly_snapshots} and thus to a stronger interaction.  In summary, due to both reasons with decreasing temperature the effective interaction and thus temperature-dependent effects become stronger. This naturally rationalizes (1) and (2).
Second, the guiding elements in the continuous model can locally translate and thus adjacent guiding elements can come together to accelerate the local assembly process. In contrast, in the lattice model the guiding elements are strictly localized. This may be one reason for the observation (3). Also other differences should be mentioned. For example the actual growth of the assembly in the continuous case may be slowed down due to the finite diffusive transport of particles to the active region. In contrast, in the lattice model the availability of additional particles is just governed by the chemical potential.

Finally, we note that the reduction in cmc upon addition of guiding elements can be further enhanced if the interaction between guiding elements and surfactant molecules is very strong. Based on our theoretical understanding it would be possible to predict the reduction in cmc in dependence on that interaction strength.

\section*{Acknowledgement}

We gratefully acknowledge financial support from the DFG (HE 2570/3-1 / TRR 61 / SFB 858) and helpful discussions with D. Liu.

\bibliography{bibliography}

% \widetext
% \clearpage
% \section{Supplementary information}\label{sec:si}

\end{document}